\documentclass[aps,pr,reprint,groupedaddress]{revtex4-2}
\input epsf.sty
\usepackage{graphicx,amsmath,amssymb,xcolor,hyperref,ulem}
\hypersetup{colorlinks=true,linkcolor=blue}
\definecolor{red}{cmyk}{0,0.8,1,0}
\definecolor{blue}{cmyk}{1,0.5,0,0}
\definecolor{green}{cmyk}{0.97,0,0.75,0}
\definecolor{cyan}{cmyk}{0.8,0,0,0}
\definecolor{magenta}{cmyk}{0.1,0.7,0,0}
\definecolor{yellow}{cmyk}{0.1,0.05,0.9,0}
\definecolor{orange}{cmyk}{0,0.5,1,0}
 
\def\lromn#1{\uppercase\expandafter{\romannumeral#1}}

\usepackage{etoolbox,xcolor}

\makeatletter

\pretocmd\frontmatter@thefootnote{\color{black}}{}{}

\makeatother

\begin{document}

\title{
A common origin of two accelerating universes: inflation and dark energy
}

\author{Kunio Kaneta}
\email{kaneta@ed.niigata-u.ac.jp}
\affiliation{
Faculty of Education, Niigata University, \\
Niigata 050-2181, Japan}

\author{Kin-ya Oda}
\email{odakin@lab.twcu.ac.jp}
\affiliation{Department of Mathematics, 
Tokyo Woman's Christian University, \\
Tokyo 167-8585, Japan}

\author{Motohiko Yoshimura}
\email{yoshim@okayama-u.ac.jp}
\affiliation{Research Institute for Interdisciplinary Science,
Okayama University \\
Tsushima-naka 3-1-1 Kita-ku Okayama
700-8530 Japan}

\date{\today}

\begin{abstract}

We develop a quantum theory of inflaton  and its decay product
of various gauge boson pairs to investigate the preheating towards
 thermalized universe.
The inflaton decay into gauge-boson pairs is shown to be inevitably accompanied by
tachyon-mass-like correction to inflation potential
 that ultimately leads to an inflaton escape out of trapped
local potential minimum towards the field infinity.
This gives rise to a conversion mechanism of early inflationary acceleration to
a quintessence dark energy acceleration at late stages of cosmic evolution.
The success of the escape depends on  how standard particles are incorporated
into a scheme of extended Jordan-Brans-Dicke gravity.
New types of  super-radiance mechanism that enhance the ending of preheating
are identified and compared with the Dicke model.

\end{abstract}

%\keywords{}

\maketitle

\section
{Introduction}

The prime objective of cosmology is to clarify how hot big bang universe started
and how the initial universe developed to the present state we observe now.
Issues related to this objective include inflation, creation of matter and radiation,
baryon asymmetry, nucleo-synthesis, dark matter, and dark energy.
We report in the present paper our recent work related to some of these problems.

The inflationary scenario solved the outstanding problems of flatness and
horizon that face the early universe by introducing a scalar field with its approximate
cosmological constant behavior driving the exponential expansion
 \cite{inflation}, \cite{inflation 2}, \cite{cosmology textbooks}.
Inflation  however leaves an empty space devoid of radiation and matter.
It is believed that field oscillation at the end of inflation
leads to copious particle production.
The popular mechanism of particle production is based on the
parametric amplification such as described by the Mathieu
(a special class of Floquet-type) equation
of periodic coefficient due to field oscillation.
Despite of many interesting achievements 
along this direction \cite{dk} --\cite{preheating review},
the outcome still does not ensure that hot big bang universe emerges necessarily.
This is one of the most important problems we  face in early cosmology.

Another important problem  in cosmology is relation
of inflation at an early epoch and dark energy at late epochs.
We explore possibilities of these two accelerating universes arising
from the same origin.
In \cite{koy 24-1} we presented one of these possibilities using
two-component scalar fields.
In the present work we give a more attractive possibility based on
a single inflaton.

It is inevitable that 
inflation and the late-time acceleration is caused by a scalar field or scalar fields.
It is desirable to introduce the scalar inflaton from some definite theoretical principle.
We pay special attention to   a class of extended
Jordan-Brans-Dicke (eJBD) \cite{jbd} scalar-tensor gravity.
It is found that our extension \cite{koy 23} 
of the original JBD and massive JBD theory \cite{ejbd}
brings about new changes to the original JBD gravity, and the problem of how
standard or grand unified particle physics is incorporated
becomes crucial  in these scalar-tensor gravity. 

We have further developed, with regard to the preheating stage of
particle production, a new idea \cite{koy 24-2} of  super-radiant
particle production of inflaton decay.
Super-radiant decay arises in a phase coherent region,
speeding up the  transfer of inflaton energy to those of produced particles.
In the present work we provide a quantum theory treating
 both inflaton and produced particle on an equal footing,
thereby unifying the two stages of parametrically amplified particle production
and super-radiant ending of preheating.
This approach bypasses a complicated mode summation method
often necessary in parametric amplification schemes of the Mathieu system
developed
in \cite{kls}, \cite{my 95 and fkyy}.

The most important result of the present quantum formulation
is that a smooth inflaton exit out of a trapped local potential minimum 
to the field infinity  is realized,
 and a quintessence type \cite{quintessence}, \cite{cosmology textbooks}
of accelerating universe emerges.
This is due to our new discovery of tachyon instability associated with particle production,
which makes the exit out of the trapped region  successful for a single inflaton field.
We thus solve the problem of relating two accelerating universes.

We avoid in this work to use the terminology of reheating often used in the literature,
since it is neither clear nor necessary that there has been a stage of
heated universe  prior to inflation.

Throughout this paper we use the unit of $\hbar = c = k_B = 1$
except that we recover $\hbar$ in a part when we discuss the WKB approximation.

\vspace{0.5cm}
Contents of the rest are organized as follows.

\begin{itemize}
\item
[\lromn2] Extended Jordan-Brans-Dicke theory: How particle physics is incorporated

A. \; Coupling to ordinary matter in lagrangian

B. \; Trace anomaly and inflaton decay to gauge boson pairs

C.\; Comparison with other popular schemes of matter coupling
\item[\lromn3]
Quantum theory of preheating

A. \; Quantum formalism

B. \; Perturbation theory to the second  orders

C. \; Tachyon instability associated with particle-pair production

D. \; {Results of higher order perturbation}
\item[\lromn4]
Super-radiant stage of preheating

A. \; Dicke model and its identification to our quantum theory

B.\; Macro-coherent amplification
\item[\lromn5]
Transition from inflationary to dark energy stages of acceleration

A. \; Inflaton escape

B. \; 
Epochs of thermalization and potential restoration
\item[\lromn6]
Summary and outlook
\item[\lromn7]
Appendix: Relevance of trace anomaly
\end{itemize}

\section
{Extended Jordan-Brans-Dicke theory: How particle physics is incorporated}
\subsection
{Coupling to ordinary matter in lagrangian}

The lagrangian density of our extended Jordan-Brans-Dicke (eJBD)
 theory is derived by a Weyl transformation from the Jordan metric frame
to the Einstein metric frame. 
The cosmological constant problem present in the Jordan frame
is absent in the Einstein frame if eJBD field decreases exponentially towards the field infinity.
Besides the eJBD gravity part,
it is important to specify how our eJBD field denoted by $\chi$ couples to standard particle fields
(gauge fields, quarks, leptons, and Higgs bosons). Our scheme uses the lagrangian density 
in the Einstein frame,
\begin{eqnarray}
&&
{\cal L} = - \frac{M_{\rm P}^2}{2} R
+ \frac{1}{2} (\partial \chi)^2 - V_{\chi} - \frac{1}{4} F_{\mu\nu}F^{\mu\nu} 
\nonumber \\ &&
\hspace*{-0.5cm}
- \bar{\psi}  \left(  \gamma   ( i \partial - e^{-\gamma_g \chi/M_{\rm P} } g A) 
- e^{- \gamma_y \chi/M_{\rm P} }  y_{\psi}\, H  \right) \psi
+ \cdots
\,.
\label {standard lagrangian}
\end{eqnarray}
Conformal factors $e^{-\gamma_i \chi/M_{\rm P}}$ appear in the terms
for which neither gauge invariance nor conformal symmetry is at work.
Two dimensionless parameters $\gamma_g, \gamma_y$ 
in this formula characterize the strength of
inflaton coupling to matter,  with 
$M_{\rm P} = 1/\sqrt{8\pi G} \sim 2.4 \times 10^{18}\,$GeV.
We shall compare later this matter coupling to more popular ones 
akin to the original  JBD gravity.

The $\chi$ inflaton potential $V_{\chi}$ is taken as
\begin{eqnarray}
&&
V_{\chi} = V_0 \, (\frac{\chi}{M_{\rm P}})^2 \,e^{ - \gamma_{\chi} \chi/M_{\rm P}}
\,, 
\\ &&
V_0 > 0
\,, \hspace{0.3cm}
\gamma_{\chi} > 0
\,.
\nonumber 
\end{eqnarray}
This potential has two extrema of $\chi = (0, 2/\gamma_{\chi})M_{\rm P}$:
the minimum is at $\chi=0$, with its potential value $0$.
The potential around the local minimum is approximated by a harmonic oscillator (HO);
\begin{eqnarray}
&&
V_{\rm HO} = \frac{V_0}{M_{\rm P}^2 }\, \chi^2
 \,.
\end{eqnarray}
The inflaton mass $m_{\chi}$ at the local minimum is
$ \sqrt{2 V_0/M_{\rm P}^2}$, estimated to be around $2 \times 10^{12}\, $GeV
 from our previous studies \cite{koy 24-1}.

\begin{figure*}[htbp]
 \begin{center}
\includegraphics[width=1.0\textwidth]{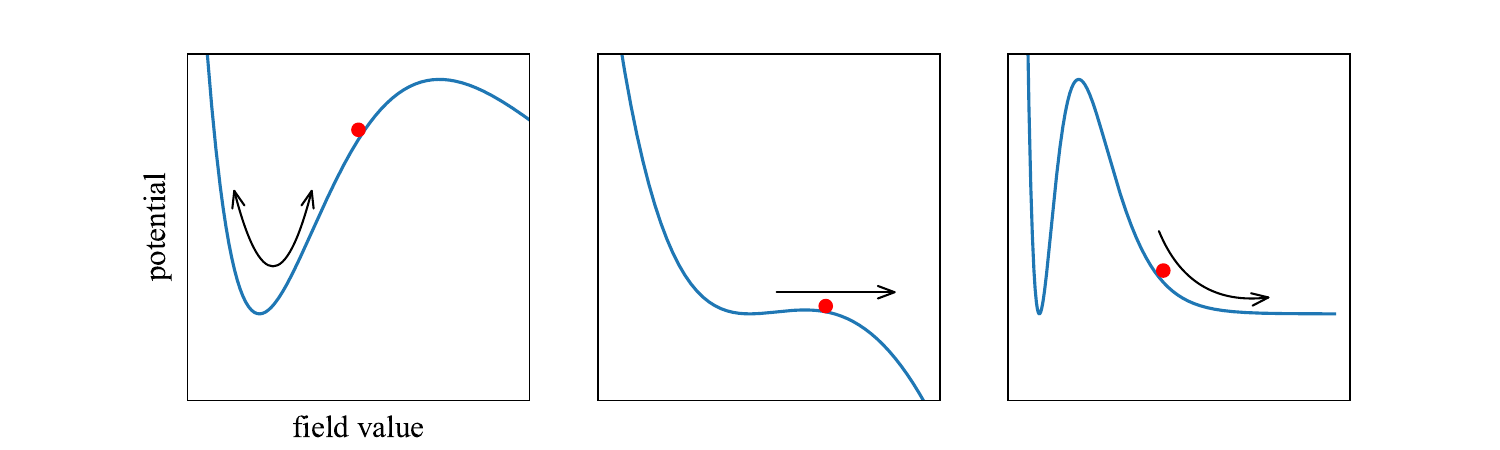}
   \caption{
Time evolution of global potential shapes:
inflaton oscillating epoch at the left, de-stabilized epoch in the middle,
and late-time epoch after the potential recovery.
}
   \label {original and corrected potential}
 \end{center} 
\end{figure*}

The potential maximum is at $\chi = 2M_{\rm P}/\gamma_{\chi}$,
with its maximum value, 
\begin{eqnarray}
&&
V_{\rm max} = \frac{4 e^{-2}\, V_0}{\gamma_{\chi}^2}
= 0.541 \frac{ V_0 }{\gamma_{\chi}^2}
\,.
\end{eqnarray}
The potential shape $V_{\chi}/V_0$ is illustrated in 
Fig(\ref{original and corrected potential})
which describes harmonic motion around the field origin for inflation, and
the quintessential motion towards the field infinity $\chi \rightarrow + \infty$.

It was shown in \cite{koy 24-1} that a variant of similar $\chi$ potential
leads to a successful inflation consistent with the observed spectral index and
limit of tensor to scalar ratio.
A further simplified model here also does the successful job.
What is important in the potential choice is existence of harmonic oscillator
part surrounded by a barrier decreasing towards the field infinity,
and detailed potential form is irrelevant to simultaneously
realize inflation and the dark energy.
Still, it is amusing to note that the potential we propose here is a simple product
of popular chaotic and quintessence potentials, 
\cite{inflation}, \cite{cosmology textbooks}, \cite{quintessence}.
Cosmological constant present in the Jordan metric frame becomes
exponentially suppressed in the Einstein frame, thus easing the cosmological 
constant problem.

\begin{figure*}[htbp]
 \begin{center}
\includegraphics[width=0.6\textwidth]{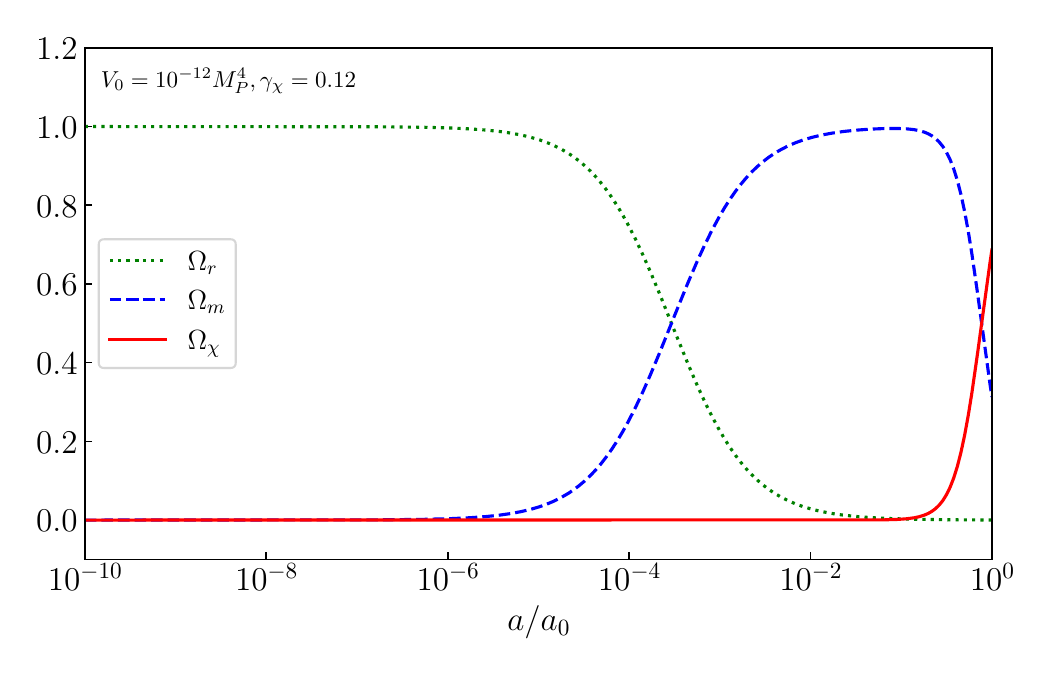}
   \caption{
Late-time evolution since nucleo-synthesis till the present time.
Assumed parameters are $\Omega_\chi(a_0)=0.687, w_\chi(a_0)=-0.998$.
}
   \label {late-time evolution}
 \end{center} 
\end{figure*}

\subsection
{Trace anomaly and inflaton decay to gauge boson pairs}

Our eJBD field evolves according to the wave equation under the flat
Friedman-Robertson-Walker metric,
\begin{eqnarray}
&&
\partial^2 \chi + \partial_{\chi} V_{\chi} = \bar{T}
\,, 
\\ &&
\bar{T} = g_{\mu\nu}\, 2 \frac{\delta (\sqrt{g} {\cal L} )}{\delta g_{\mu \nu } } 
= \bar{T}_m + 
\left({\rm graviton\; and\; } \chi\; {\rm parts} \right)
\,,
\label {ejbd field eq}
\end{eqnarray}
with $\bar{T}$ the trace of energy-momentum tensor including the small gravitational coupling of order
$ 1/M_{\rm P}$ in the expansion, as verified by the expansion, 
$e^{-\gamma_g \chi/M_{\rm P}} \sim 1 -\gamma_g \chi/M_{\rm P} $.
The first term  $\bar{T}_m$ in (\ref{ejbd field eq}) in the right hand side 
is the contribution from standard matter fields.
Right after inflation and when inflaton starts oscillation, the gauge symmetry is spontaneously broken since there is no temperature effect restoring
the symmetry,
but time dependent inflaton mass is of the order, $M_{\rm P}$,
hence all standard particle masses may be regarded as nearly massless.
This  makes $\bar{T}_m$ nearly vanishing.

There is however an important quantum correction of  the trace anomaly \cite{ps}. 
The trace-anomaly related contribution gives  an effective decay amplitude of
$\chi \rightarrow A_1 A_2$ with  $A_i \,, i=1,2$ two massless gauge fields,
as fully explained in Appendix below. The amplitude is
\begin{eqnarray}
&&
\gamma_J \, \frac{ \chi}{M_{\rm P}}  \sum_A\,
\frac{b_A \alpha_A^2}{8\pi} \, (F^{ A\, (1)} )^{\rho\sigma} \,F^{ A\, (2)} _{\rho\sigma} 
\,.
\label {trace anomaly of gauge-pair}
\end{eqnarray}
 $\alpha_A = g_A^2/4\pi$ are various gauge couplings, and $b_A$ is the coefficient of
renormalization group.
There is essentially no observational constraint on the new conformal factor
$\gamma_J$ here, and this decay gives a promising candidate for
the instability we work out.

In grand unified theories there are other important contributions
whose implications shall be discussed elsewhere.

In analogy to QED (Quantum Electro Dynamics) the squared field strength may be
expressed in terms of electric and magnetic type gauge fields of $A$ types;
\begin{eqnarray}
&&
\frac{1}{2} (F^{A\,(1)}_{\rho\sigma}) (F^{A\,(2)})^{\rho\sigma}
= \left( \vec{E}^{A\, (1)}\cdot \vec{E}^{A\, (2)}  
- \vec{B}^{A\, (1)}\cdot \vec{B}^{A\, (2)}  \right)
\nonumber \\ &&
\\ &&
+ ({\rm cubic \; terms\; for\; non-Abelian\; gauge\; fields})
\,.
\nonumber
\end{eqnarray}
Estimate in SO(10) grand unification gives 
$\sum_A \frac{b_A \alpha_A}{16\pi} \sim -0.014$ \cite{koy 24-2}.
We shall often omit gauge field species $i$ and its summation $\sum_A$
for notational simplicity.

Inflaton decay rate $\Gamma_B$ in perturbation theory is readily calculable, assuming
no time variation of various constants, to give
\begin{eqnarray}
&&
\Gamma_B = 
\gamma_J^2\,(\sum_A \frac{b_A \alpha_A}{8\pi})^2
\, \frac{ m_{\chi}^3}{16 \pi^2 M_{\rm P}^2}
\,.
\end{eqnarray}
We shall show later that this rate is dramatically enhanced
leading to almost instantaneous decay at the super-radiant phase
of preheating.

Since inflatons are at rest in all stages of cosmic evolution, 
this decay is characterized by back-to-back pair production
of momentum configuration, $\vec{k}\,, - \vec{k}$ with its magnitude
fixed by $|\vec{k}| \equiv k = m_{\chi}/2$.
This is important to create macro-coherence at super-radiance stage in which
the phase coherence region is not limited by the wavelength of light unlike in
the original Dicke model \cite{dicke}--\cite{sr textbook}.
This is an enormous advantage to realize a rapid termination of preheating,
since the use of decay rate of the Born approximation
leads to slow decay accompanied by the difficulty of thermalization.

\subsection
{Comparison with other popular schemes of matter coupling}

It would be instructive to compare our eJBD matter-coupling 
to other more popular ones  based on the original JBD idea \cite{jbd}.
A popular scheme introduced in \cite{def} uses the matter coupling
in the Einstein metric frame; taking $\alpha_0, \beta_0$ as constants,
\begin{eqnarray}
&&
{\cal L}_m = {\cal L}_{st} ( \psi_m; A^2 g_{\mu\nu})
\,, 
\\ &&
\hspace*{-0.5cm}
A^2 = A^2(\chi) = \exp[- \alpha_0 \frac{\chi}{\sqrt{8\pi}\, M_{\rm P} } 
+ \beta_0 (\frac{\chi}{\sqrt{8\pi}\, M_{\rm P} } )^2]
\,,
\end{eqnarray}
with $\psi_m$ representing standard fields generically.
(We have slightly changed the reference point of $\chi$ in this formula
from \cite{def}, whose dimensionless $\chi$ is not well defined, but we believe the
formula above agreeing with their results.)
The original JBD uses the linear term alone with $\beta_0 =0$.

Despite of apparent difference of this parametrization from ours,
these are closely related.
It is sufficient to inspect the contribution of anomaly term in the field equation 
(\ref{ejbd field eq}).
$A^2$ parameterization gives, in the eJBD massless limit,
\begin{eqnarray}
&&
\partial^2 \chi = -  \frac{\chi}{\sqrt{8\pi}\, M_{\rm P}} \,\alpha_0 \bar{T}_{\rm anomaly} 
+ O\left(
(\frac{\chi}{M_{\rm P}} )^2\right)
\end{eqnarray}
while ours gives to the right hand side
\begin{eqnarray}
&&
-  \frac{\chi}{ M_{\rm P} } \, \gamma_J \sum_A  \bar{T}_{\rm anomaly} + O\left(
(\frac{\chi}{M_{\rm P}} )^2\right)
\,.
\label {ejbd-matter coupling: ours}
\end{eqnarray}
Thus, to this order,
\begin{eqnarray}
&&
\alpha_0 = \sqrt{8\pi}\, 
 \gamma_J
\,, \hspace{0.3cm}
\beta_0 = 0
\,,
\end{eqnarray}
for a single gauge field.
Correspondence between two coupling schemes
 becomes more complicated if one includes contributions from
other sources of quark mass term and Higgs contribution.
But one thing is clear: the popular scheme emphasizes the contribution from
$\propto \beta_0$ term in the exponent of $A^2(\chi)$,
while ours emphasizes multiple contributions of $\alpha_0$ types,
although one can readily incorporate in our formalism $\beta_0$ type exponents.

Our coupling scheme (\ref{ejbd-matter coupling: ours}) is an extension of
 the original JBD gravity \cite{jbd} which was invented at a time 
when only QED is firmly established.
Our present scheme fully respects
 the gauge invariance of established particle physics in the present days, but
abandon the  uniform and simplest form of Weyl conformal transformation.
as stressed in \cite{koy 24-1}.
The standard model of particle physics is based on the spontaneously broken gauge symmetry,
and the theory is SU(3) $\times $ SU(2) $\times $ U(1) gauge invariant at the lagrangian level.
There are five pieces of gauge invariant lagrangian densities in the action principle;
squared gauge field-strength, fermion kinetic term, fermion Yukawa coupling,
Higgs kinetic term, and Higgs potential.
Our principle is that these five pieces are independently Weyl rescaled.
By redefining field operators according to the canonical formalism, one can have a plethora of conformal $A^2$ factors.
One of the simplest forms is the one given here.
This form emerges when higher loop corrections are incorporated in super-string theory 
\cite{damour-polyakov}.
Since it is difficult to convincingly derive the low energy limit of the super-string theory,
we shall follow a bottom-up approach using our parametrization.

\section
{Quantum theory of preheating}
\subsection
{Quantum formalism}

 Events after inflation occur independently at different
space points within the Hubble horizon, with its size given by
$H^{-1} = a/\dot{a} $, $a(t)$ being the cosmic scale factor.
Hence we may introduce independent inflaton fields,
 $b_s(t), b_s^{\dagger}(t)$, at different sites $s$.
All inflatons are assumed  at rest with zero momentum. We thus introduce their hamiltonian,
\begin{eqnarray}
&&
\chi(t) = 
\frac{1}{\sqrt{2m_{\chi}V} }\, \sum_s \, ( b_s(t) \, e^{- i m_{\chi} t} 
+ b_s^{\dagger}(t) \, e^{ i m_{\chi} t} )
\,, 
\label {q-inflaton field}
\\ &&
[ b_s(t) \,, b_{s'}^{\dagger}(t)] = \delta_{s, s'}
\,, \hspace{0.3cm}
[ b_s(t)\,, b_{s'}(t)] = 0
\,,
\\ &&
H_0(\chi) =
\int d^3x_s\, \frac{1}{2} (\, \pi_{\chi}^2 + m_{\chi}^2 \chi^2\, ) = 
m_{\chi} \sum_s \left( 
b_s^{\dagger}  b_s + \frac{1}{2} \right) 
\nonumber \\ &&
+ ({\rm oscillating\; terms})
\,.
\end{eqnarray}
Site variable $s$ in operators $b_s(t)\,, b_s^{\dagger}(t)$
are merely labels for distinction, and there is no actual coordinate $\vec{x}_s$ dependence 
due to the space translational invariance of inflaton time evolution,
at the initial time of time evolution.

We may write interaction hamiltonian that holds in the interaction picture;
\begin{eqnarray}
&&
H_{\rm int} = - c_A  \int d^3x_s\, \chi(t) (\vec{E}^2 - \vec{B}^2)
\\ &&
 = -   c_A \int \frac{V d^3k}{(2\pi)^3}\, \sum_s \, \frac{2 \omega_k }{\sqrt{m_{\chi} V }}
\nonumber \\ &&
\sum_{\rm pol}\,
\left( \vec{e}(k)\cdot \vec{e}(-k) \, b_s^{\dagger}(t) \, a_k a_{-k} \, e^{i (m_{\chi}-2 \omega_k )t} 
\right.
\nonumber \\ &&
+ \vec{e}(k)^*\cdot \vec{e}(-k)^* \, b_s(t)\, a_k^{\dagger} a_{-k}^{\dagger} \, 
e^{i(- m_{\chi} +2 \omega_k) t} 
\label {on-shell interaction}
\\ &&
+\vec{e}(k)\cdot \vec{e}(-k)  b_s(t)\, a_k a_{-k} \, e^{- i (m_{\chi}+ 2 \omega_k )t} 
\nonumber \\ &&
\left.
+ \vec{e}(k)^*\cdot \vec{e}(-k)^* b_s^{\dagger}(t) \, a_k^{\dagger} a_{-k}^{\dagger} \, 
e^{i(m_{\chi} +2 \omega_k) t} 
\right)
\,, 
\\ &&
c_A = \frac{\gamma_J \, b_A \alpha_A^2 }{8 \pi M_{\rm P}}
\,.
\label {off-shell interaction}
\end{eqnarray}
Operators $a_k, a_k^{\dagger}$ are annihilation and creation operators
of various gauge vector potentials using the radiation gauge. 
The first  two terms in (\ref{on-shell interaction})
give contributions that survive on the mass shell at $m_{\chi}=2 \omega_k$,
while the second two contributions in (\ref{off-shell interaction}) are those
from off the mass shell.
Operators, $a_{\pm k}, a_{\pm k}^{\dagger}, b_s$, are dimensionless.
We shall often consider, for simplicity, the maximum polarization sum 
$\sum_{\rm pol}\vec{e}(k)\cdot \vec{e}(-k) =2$ (happens to coincide with
a squeezed state),
which occurs when the back-to-back gauge field pair has either parallel or anti-parallel 
linear polarizations.

So far we discussed gauge boson pair coupling to eJBD $\chi$ field.
If the equally possible Higgs boson pair is adopted,
the coupling is replaced by 
$c_A = y_{\psi}^2 \gamma_y /(64\pi^2 M_{\rm P}) $
with weaker Yukawa coupling  $y_{\psi}^2/4\pi $ than $\alpha$.
We shall not purse this possibility any further in the present work.

We separate the on-shell contribution and denote it by the prime $'$,
\begin{eqnarray}
&&
H_{\rm int}' (t)
 = -   4\sqrt{2} c_A \, \int \frac{V d^3k}{(2\pi)^3}\, \frac{ \omega_k }{\sqrt{ 2m_{\chi} V}}\, \sum_s\,
\nonumber \\ &&
\left( b_s^{\dagger} \, a_k a_{-k} \, e^{i (m_{\chi}-2 \omega_k )t} 
+  b_s\, a_k^{\dagger} a_{-k}^{\dagger} \, 
e^{i(- m_{\chi} +2 \omega_k) t} 
\right)
\,.
\label {on-shell interaction 2}
\end{eqnarray}
The volume dependence  shown here is natural, since the 
positive and negative frequency components of quantum inflaton field are 
\begin{eqnarray}
&&
\chi^{(+)}(t) = \frac{1}{\sqrt{2m_{\chi}V} } \sum_s b_s(t) \, e^{- i m_{\chi} t} 
\,, 
\\ &&
\chi^{(-)}(t) = \frac{1}{\sqrt{2m_{\chi}V} }\sum_s  b_s^{\dagger}(t) \, e^{ i m_{\chi} t} 
\,,
\end{eqnarray}
with the conventional volume dependent $\sqrt{V}$ factor included.
This on-shell contribution $H_{\rm int}' (t) $ is akin to the short time average previously considered
in \cite{koy 24-2}.
The volume $V$ in cosmology may effectively be taken as the Hubble volume $H^{-3}$.
The time dependence of inflaton in the harmonic potential  simplifies to
$ \chi^{(\pm)}(t) = \chi^{(\pm)}(0)\, e^{ \mp i m_{\chi} t}$, 
while it has extra time dependence
under more general $\chi$ potentials, $V_{\chi}$.

As initial states we take states without radiation pair.
It is necessary to distinguish states of
quantum $\chi$ and radiation fields separately;
at a single site we may take excited states as well,
hence use the notation $|n_s\rangle_{\chi}\,, n_s =0, 1,2, \cdots$.
The general class of initial  states at ${\cal N}_s$ sites is a direct product;
\begin{eqnarray}
&&
|\Psi[\{n_s \}, 0] \rangle = \Pi_s \, (|n_s\rangle_{\chi} \, | 0 \rangle_{\rm rad}\, )
\,.
\end{eqnarray}
Radiation quanta are dispersed in space, and there is no need
to attach them to particular sites.

Spatially homogeneous states are however adequate for consideration
right after the end of inflation of the kinds, hence we only consider the type
of initial states,
\begin{eqnarray}
&&
|\Psi(n; 0) \rangle = \Pi_s \, ( |n\rangle_{\chi,s }  | 0 \rangle_{\rm rad}\,)
\,.
\end{eqnarray}
Dicke-type initial state we later consider is $|\Psi(1; 0) \rangle $.
We also need to consider excited states such as
$|\Psi[{0_s, 2_{s'\,} }, 2 \rangle $, 
meaning that $\chi$ is in the ground state only at site $s$,
and is in the second excited states at all other sites.
Time evolved states are not spatially homogeneous and
the distribution of $\chi$ state at different site is automatically 
described in our formalism.

States evolve by a unitary operator, and for instance 
the Dicke-type state in the interaction picture follows according to
\begin{eqnarray}
&&
|\Psi (1; 0) \rangle_i \rightarrow |\Psi (t)\rangle = U(t, t_i) |\Psi (1;0) \rangle_i 
\,, 
\\ &&
U(t, t_i) = {\cal T}\, \exp[- i \int_{t_i}^t dt'\, H'_{\rm int} (t'- t_i)]]
\,,
\end{eqnarray}
with ${\cal T}$ the time ordering.
This is not a purely quantum evolution, since we dropped the off-shell parts of
interaction hamiltonian. We may call this approximation semi-quantum theory.
It is convenient to introduce a conjugate pair of combined field operators,
\begin{eqnarray}
&&
\hspace*{-0.2cm}
Q^{(+)} (t) =  \chi^{(+)}(t) a_k^{\dagger} a_{-k}^{\dagger}
\,, \hspace{0.3cm}
Q^{(-)} (t) =  \chi^{(-)}(t) a_k a_{-k}
\,.
\nonumber \\ &&
\end{eqnarray}
Using conjugate dynamical variables $Q^{(\pm)}(t)$,
the interaction hamiltonian becomes
\begin{eqnarray}
&&
H_{\rm int}' (t) = -   4\sqrt{2} c_A \int \frac{V d^3k}{(2\pi)^3}\,\omega_k\, 
e^{2 i \omega_k t} Q^{(+)} (t) +({\rm h.c.})
\,,
\nonumber \\ &&
\label {interaction hamiltonian 5}
\end{eqnarray}
The relevant time integration is
\begin{eqnarray}
&&
- i \int_{t_i}^t dt'\, H'_{\rm int} (t'- t_i) 
\nonumber \\ &&
=
- 4 \sqrt{2} c_A \int \frac{V d^3k}{(2\pi)^3}\, \frac{\omega_k }{2 \omega_k -m_{\chi} }
\times
\nonumber \\ &&
 \left[ 
\left( \cos (m_{\chi}- 2 \omega_k )(t- t_i) - 1 \right) 
\right.
\left( Q_{\rm HO}^{(+)} (0)  - Q_{\rm HO}^{(-)} (0)  \right) 
\nonumber \\ &&
\hspace*{-0.3cm}
\left.
+ i \sin (m_{\chi}- 2 \omega_k )(t- t_i) 
\left( Q_{\rm HO}^{(+)} (0)  + Q_{\rm HO}^{(-)} (0)  \right) 
\right]
\,. 
\label {time-integrated hamiltonian}
\end{eqnarray}
 for the pure HO case.

For convenience
we introduce the phase space factor relevant in cosmology,
\begin{eqnarray}
&&
\Omega_k \equiv \int \frac{V d^3k}{(2\pi)^3} = \frac{ H^{-3}} { 8\pi^2} m_{\chi}^2 | 
\Delta \omega_k |
\,, 
\\ &&
\hspace*{1cm}
\Delta \omega_k = \omega_k - \frac{m_{\chi}}{2}
\,,
\end{eqnarray}
with $1/H$ the Hubble length.
The quantity $\Delta \omega_k$ is taken as a fraction of $m_{\chi}/2$.
The real part of operator coefficients in time-integrated hamiltonian
 behaves like $\propto 2 \Delta \omega_k \, (t-t_i)^2 $, as $t-t_i \rightarrow 0$,
while its imaginary part behaves like $\propto  i \omega_k (t-t_i)$.
These behaviors are typical of quantum fluctuation and dissipation.

\subsection
{Perturbation theory to the second order}

Let us systematically work out more details of gauge boson pair production amplitude
and expectation values in initial states.

To the first order perturbation, 
there is only type of decay amplitude,
\begin{eqnarray}
&&
- i \, \langle \Psi(0;1) | \int_0^t dt' H_{\rm int}'(t') | \Psi(1; 0) \rangle
\,,
\end{eqnarray}
whose square gives decay rate after a phase space integration.

To the second order of interaction, there are two types of decay amplitudes,
the difference due to the involved number of sites, one or two.
They are
\begin{eqnarray}
&&
(1)\;
{\cal A}_2 = 
\nonumber \\ &&
\hspace*{-0.5cm}
- \frac{1}{2} \langle \Psi[{0_{s1}, 0_{s2},2_{s'' }}, 2]  | {\cal T}\, \left( 
\int_0^t dt' H_{\rm int}'(t')  \right)^2 | \Psi(1; 0) \rangle
\,,
\\ &&
(2)\; 
{\cal A}_2' = 
\nonumber \\ &&
- \frac{1}{2} \langle \Psi[{0_s, 2_{s' }}, 2] | {\cal T}\, 
\left( \int_0^t dt' H_{\rm int}'(t') \right)^2 | \Psi(2; 0) \rangle
\,.
\end{eqnarray}
Site occupancy is different in the two cases:
there are two sites in the ground state in (1), while there is only one
in (2).
This is due to that the initial states are different.
Decay amplitudes and decay rates are different in the two cases,
but what might be confusing is that to the leading $(\Delta \omega_k\, t)^2$
expansion results are the same;
\begin{eqnarray}
&&
\hspace*{-0.5cm}
(1)\;  {\cal A}_2 =
\frac{\tilde{\chi}^2}{2} \, (1- e^{2i \Delta \omega_k\, t })^2
= - 2 \tilde{\chi}^2 (\Delta \omega_k\, t )^2 + \cdots
\,,
\\ &&
\hspace*{-0.5cm}
\tilde{\chi} = 
2 \sqrt{2} c_A \chi^{(+)}\, \frac{m_{\chi} \Omega_k} {2 \Delta \omega_k }
= \frac{ \sqrt{2} }{16 \pi^2 }(\frac{3 }{2 })^3 c_A \chi^{(+)}\, (m_{\chi} t)^3
\,,
\label {basic coupling in perturbation th}
\\ &&
\hspace*{-0.6cm}
(2)\; {\cal A}_2' = - \frac{\tilde{\chi}^2}{2} \, |1- e^{2i \Delta \omega_k\, t })|^2
= - 2 \tilde{\chi}^2 (\Delta \omega_k\, t )^2 + \cdots
\,.
\end{eqnarray}
The difference of $  {\cal A}_2$ and $ {\cal A}_2' $
appears in the neglected dotted terms.

The decay amplitude ${\cal A}_2$ in case (1) corresponds to the mode-summed probability
amplitude in the Mathieu chart in past approaches \cite{kls}, \cite{my 95 and fkyy},
and exponentially grows in time.
The decay amplitude is however dramatically changed when
higher order effects are included, and grows much faster
beyond critical time, as is made clear below.

It becomes important to clarify time dependence of field and
mass parameter $m_{\chi}$ that appear in the basic coupling $\tilde{\chi}$
in (\ref{basic coupling in perturbation th}).
Time variation of field $\chi^{(\pm)} = \chi/2$ follows 
the law in the matter-dominated universe, 
$\rho_{\chi} (t) \propto a^{-3} \propto t^{-2}$, from
which one derives the behavior of product, $m_{\chi} \chi \propto 1/t$.
This relation is collaborated with
HO motion governed by the equation including the Hubble friction,
\begin{eqnarray}
&&
\ddot{\chi} + 3 H \dot{\chi} + m_{\chi}^2 \chi = 0
\,, \hspace{0.3cm}
H= \frac{2}{3t}
\,.
\label {ho field eq with hubble}
\end{eqnarray}
The differential equation is solved in terms of Bessel type functions, and its
long time limit is given by
\begin{eqnarray}
&&
\rho_{\chi} = m_{\chi}^2 \chi^2 \rightarrow (m_{\chi} \chi_0)^2 (\frac{t_0}{t})^2
\propto a^{-3}(t)
\,.
\end{eqnarray}
Assuming a constant $m_{\chi}$, we arrive at $\chi(t) \propto 1/t$.
$t_0$ is the onset time of inflaton oscillation.

The basic coupling is then expressed by
\begin{eqnarray}
&&
\tilde{\chi} = \frac{ 27 \sqrt{2} }{ (16 \pi )^2} c_A \chi (m_{\chi} t)^3 \frac{\chi}{M_{\rm P}}
\nonumber \\ &&
= \frac{3 \sqrt{6} }{32 \pi^2 } c_A M_{\rm P}\, (\frac{z_0}{z})^{9/2} \frac{\chi}{M_{\rm P}}
\,,
\end{eqnarray}
with $3 \sqrt{6}/( 32 \pi^2) = 0.0233$.
$z_0$ here is the redshift factor at $t=t_0$.
We have used as a time variable the  redshift factor $z$.
Time variation in shorter time scales are given by
oscillating functions in terms of $\Delta \omega_k t =
m_{\chi} t \times (\Delta \omega_k/m_{\chi})$ where
we take $\Delta \omega_k/m_{\chi} = O(0.1 \sim 0.01)$.

We should have expected uncontrollable divergence caused by
the interaction hamiltonian $\chi F_{\mu\nu}F^{\mu\nu}$
which is non-renormalizable operator,
but the gauge invariance makes the self-energy controllable.

\subsection
{Tachyon instability 
associated with particle-pair production}

\begin{figure*}[htbp]
 \begin{center}
\includegraphics[width=0.6\textwidth]{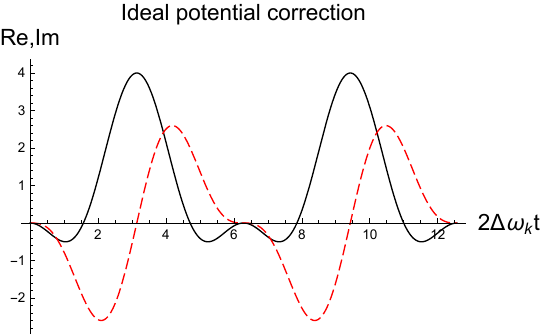}
   \caption{
Ideal form of potential correction plotted against $2\Delta \omega_k\, t$:
the real part given by (\ref{ideal potential correction real}) in solid black
and the imaginary part given by (\ref{ideal potential correction imaginary}) in
dashed orange. For simplicity $\tilde{\chi}^2/2 =1 $ is assumed.
When the imaginary part in orange is positive and the real part in black is negative,
the strong parametric resonance effect is at work in instability bands.
}
   \label {ideal potential correction}
 \end{center} 
\end{figure*}

What is  important equally to the decay amplitude
is the expectation value in the initial state.
This describes quantum fluctuation, and its interpretation
gives $\chi$ potential correction related to particle production.
The relation of the two constitutes an important part of fluctuation-dissipation theorem.
The expectation value in the Dicke initial state is given by
\begin{eqnarray}
&&
- \frac{1}{2} \, \langle \Psi(1; 0) |{\cal T} \left( \int_0^t dt' H_{\rm int}'(t') \right)^2
| \Psi(1; 0) \rangle 
=  {\cal E}_2 
\,,
\\ &&
\Re {\cal E}_2 = - \frac{ \tilde{\chi}^2 }{2} \left( 1- 2 \cos (2 \Delta \omega_k\, t) 
+ \cos ( 4 \Delta \omega_k\, t) \right) 
\,,
\label {ideal potential correction real}
\\ &&
\Im {\cal E}_2 
= - \frac{ \tilde{\chi}^2 }{2} \left( - 2 \sin (2 \Delta \omega_k\, t) 
+ \sin ( 4 \Delta \omega_k\, t) \right)
\,,
\label {ideal potential correction imaginary}
\end{eqnarray}
to the second order of interaction.
We plot the real and the imaginary parts 
in Fig(\ref{ideal potential correction}).
The positive imaginary part is a kind of anti-dissipation,
and may appear in the parametric amplification.
The negative real part that starts at $2 \Delta \omega_k t = 3\pi/2$
is of particular interest: it is correlated with decreasing positive
imaginary part.
The positive imaginary part indicates the growing pair production.
In Section \lromn5 C we shall show that the escape probability
of eJBD field out of trapped potential region is greatly enhanced.

The quantum fluctuation given by the real part
 is  $\chi$ field dependent; $\propto \chi^2$.
It is thus natural to interpret this as a $\chi$ mass correction,
hence we equate the $t\rightarrow 0$ limit of real part to
\begin{eqnarray}
&& 
\int d^4x \, \delta {\cal V}_{\chi} = \frac{\delta m_{\chi}^2}{2} \chi^2 t \, (H^{-1})^3
\,.
\end{eqnarray}
This interpretation leads to an identification of mass correction,
\begin{eqnarray}
&&
\delta m_{\chi}^2 
= - c\, (c_A M_{\rm P})^2 \, M_{\rm P}^2  (\frac{z_0 }{z })^3 \,,
\label {tachyonic mass correction}
\\ &&
c = \frac{160}{3} (\frac{ 2}{ 3} )^3 \frac{(27 \sqrt{2})^2}{(16 \pi)^4 } = 3.61 \times 10^{-3}
\,.
\end{eqnarray}
We have proved that a tachyon mass with $\delta m_{\chi}^2 < 0$
emerges  to the lowest second order of interaction.
The result of negative squared mass correction implies
that  the tachyon instability near the field origin occurs.

Let us study whether this instability occurs with a reasonable time
range of $t_0$ to the termalization time $t_R$,
$t_0 \sim t_R = O(4 \times 10^3 ) t_0$.
The tachyon instability occurs at $\delta m_{\chi}^2 + m_{\chi}^2 =0$
with $m_{\chi}$ the HO mass.
This condition is equivalent to
\begin{eqnarray}
&&
- c\, (c_A M_{\rm P})^2 \,\frac{ M_{\rm P}^2}{m_{\chi} ^2} (\frac{z_0 }{z })^3 
 + 1 = 0
\,,
\end{eqnarray}
One may thus derive the red shift factor of instability time $t_{\rm tac}$,
\begin{eqnarray}
&&
z_{\rm tac} = c^{1/3} (c_A M_{\rm P})^{2/3} (\frac{M_{\rm P} }{m_{\chi} })^{2/3}  z_0
\,,
\end{eqnarray}
with $c^{1/3} = 0.153$.
For a choice of  $\gamma_J = 0.01 \sim 0.1$ in the formula
$c_A M_{\rm P} =  \gamma_J \alpha/16\pi$,
$z_{\rm tac}/z_0$  falls in its acceptable
  range of $2 \times 10^{-5} \sim 0.9 \times 10^{-4}$,
assuming $m_{\chi} = M_{\rm P}$ (smaller values than this are favored).

The tachyon instability indicated by the negative quantity 
in (\ref{tachyonic mass correction})
 is one of our major findings in the present quantum theory of preheating, and
opens an interesting possibility towards the transition 
from the early inflationary to the late dark energy acceleration 
in eJBD gravity of a single field.

One can determine $|\Delta \omega_k|$ (but not its sign) by making
an alternative calculation in the second order perturbation;
the Feynman diagram technique.
The diagram in the second order
contains a kind of polarization scalar $\Pi_s (k_1, k_2) $
made of one-loop  gauge boson pair propagation;
\begin{eqnarray}
&&
\Pi_s (k_1, k_2) = \int d^4 x\, e^{i (k_1 + k_2)\cdot x }
\nonumber \\ &&
\hspace*{-0.3cm}
\langle 0 | T\left[ \left( F^{\rho \sigma}(x) 
F_{\rho \sigma}(x) \right)
\left( F^{\mu \nu}(0)  
F_{\mu \nu}(0) \right)
\right] | 0\rangle 
\,,
\end{eqnarray}
with the field strength 
$ F_{\mu \nu}(x) = \partial_{\mu} A_{\nu} 
- \partial_{\nu} A_{\mu}$
related to the vector potential $A_{\mu}(x)$.
Using the Feynman gauge propagator of vector potential, we  derive
\begin{eqnarray}
&&
\Pi_s (k_1, k_2) = 6 \left( (k_1+k_2)^2 \right)^2
\,, 
\\ &&
k_1 = (\omega_k, \vec{k})
\,, \hspace{0.3cm}
k_2 =  (\omega_k, -\vec{k})
\,.
\end{eqnarray}
Thus, $ \Pi_s (k_1, k_2) = 6 m_{\chi}^4$, since
$\omega_k = m_{\chi}/2$ due to the energy conservation.

This simplification brings about a correspondence to the previous formula,
\begin{eqnarray}
&&
\frac{ 6 m_{\chi}^4}{ q_0^2 - \vec{q}^2 - m_{\chi}^2 + i 0^+}
\hspace{0.3cm} (q_0 = 2 \omega_k\,, \vec{q}=0)
\\ &&
\Leftrightarrow 
\frac{\Omega_k  2 m_{\chi}}{ V } \frac{1}{2\omega_k ( 2\omega_k - m_{\chi} + i 0^+) }
\, F
\,,
\\ &&
F = F(\Delta \omega_k\, t) =
\left( 1 - e^{- 2 i \Delta \omega_k \, t }  - 2 i \Delta \omega_k \, t \right)
\,.
\end{eqnarray}
The absolute value of the function $F$ here behaves as $(\Delta \omega_k t)^{3/2} $
for large times, and we conclude from this
\begin{eqnarray}
&&
|\Delta \omega_k |  = c \, M_{\rm P}(\frac{m_{\chi} }{M_{\rm P} })^{2/5 } 
 (\frac{ z}{ z_0} )^{ 9/10}
\,, 
\\ &&
 c=  (24 \pi^2)^{2/5} (\frac{\sqrt{3} }{ 2} )^{ 3/5} \sim 8.17
\\ &&
\Omega_k  \sim 0.493 \,  (\frac{m_{\chi} }{M_{\rm P} } )^{12/5 } 
 (\frac{ z_0}{ z} )^{ 18/5}
\,.
\end{eqnarray}
The value
$|\Delta \omega_k |/m_{\chi} =  1 \sim 0.1 $ corresponds to %\\
$ z/z_0 = ( 0.097 \sim 0.0075)\, (m_{\chi}/M_{\rm P})^{2/3}$.
In numerical computations that follow we shall take $ \Delta m_k = 0.1 \,m_{\chi}$
as a standard reference.

\subsection
{Results of higher order perturbation}

It is both necessary in later discussions and also illuminating for
a deeper understanding of what occurs in our system
to work out higher orders beyond the second.

The $n-$th order perturbation theory gives the expectation value in
the Dicke type initial states,
\begin{eqnarray}
&&
\hspace*{-0.5cm}
{\cal E}_n = \frac{(-i)^n }{n! } \,\langle \Psi(1; 0) | 
 {\cal T} \left( \int_{0}^{t} dt'\, H'_{\rm int}(t') \right)^n | \Psi(1; 0) \rangle
\,.
\end{eqnarray}
For this quantity to be non-vanishing, $n$ must be even positive integer,
since the final  state $\langle \, {\rm final} \, |$ goes back to the initial state 
$\langle \, {\rm initial} \,|$.
To the second order, this happens to be identical to the decay amplitude,
thus ${\cal E}_2 = {\cal A}_2$.
The relation of expectation value to the potential correction is
\begin{eqnarray}
&&
 \langle \Psi(1; 0)| U(t,0) | \Psi(1; 0)  \rangle = 
\exp[ -i t H^{-3} \langle \delta V_{\chi} \rangle]
\,.
\label {potential expectation-value rel}
\end{eqnarray}

Decay and excitation at the same site such as
$|1_s\rangle_{\chi} \rightarrow  |0_s\rangle_{\chi} \rightarrow |1_s\rangle_{\chi}$ 
is unlikely to occur due to the diffusion of produced pairs deep into the
momentum phase space.
We thus invoke an assumption that the time ordered product
contains amplitudes of $n$ pair production followed by $n$ pair annihilation
 between sandwiched initial states, 
giving  non-vanishing contributions in even orders alone.
We thus have, in the HO potential region,
\begin{eqnarray}
&&
\hspace*{-0.5cm}
_{\rm rad}\!\langle n | _{\chi}\! \Pi_s \langle 0_s| 
\left( Q^{(+)}_{\rm HO}(0) \right)^n  \Pi_s |1_s \rangle_{\chi} | 0 \rangle_{\rm rad}
 = n! \, (\chi^{(+) })^n
\,.
\label {n-th order formula}
\end{eqnarray}
Decays occur at $n$ sites out of entire $N_s$ sites.
It is reasonable to take for $\chi^{(+)} $ the half of field value,
\begin{eqnarray}
&&
\chi^{(+)} = \frac{1}{\sqrt{2 m_{\chi}V}}\, _{\chi}\!\langle 0_s| b_s | 1_s \rangle_{\chi}
= \frac{\chi}{2}
\,.
\end{eqnarray}

Since $n\, Q^{(+)}_{\rm HO}$  operators in (\ref{n-th order formula})
are independent at different sites, the relevant pair production
is restricted to $n=2$ multiplied by the involved site number.
This  gives the factor to the expectation value,
\begin{eqnarray}
&&
{\cal E}_{2n} = ({\cal E}_{2})^n = \tilde{\chi}^{2n}
\left(e^{- 2i \Delta \omega_k t} -1 + 2i \Delta \omega_k t \right)^n
\,.
\end{eqnarray}
The summation over all $2n$ orders give a simple geometric series,
\begin{eqnarray}
&&
\sum_n {\cal E}_{2n} = \frac{1}{ 1 + {\cal E}_{2}}
\,.
\end{eqnarray}
We interpret this as the potential correction $\langle \delta V_{\chi} \rangle $
using the relation,
\begin{eqnarray}
&&
\exp[- i \langle \delta V_{\chi} \rangle t H^{-3}] = \frac{1}{ 1 + {\cal E}_{2}}
\,.
\end{eqnarray}
This gives
\begin{eqnarray}
&& 
\langle \delta V_{\chi} \rangle = - i
  \frac{1}{6} M_{\rm P}^4 (\frac{z }{ z_0})^6 \ln (1 + {\cal E}_{2} )
\,.
\end{eqnarray}

From the convergence of the series, one must impose a condition,
\begin{eqnarray}
&&
|  {\cal E}_{2}| =\tilde{\chi}^2\, \left[
2 (1 - \cos 2\Delta \omega_k t) - 4 \Delta \omega_k t
\sin 2\Delta \omega_k t  
\right.
\nonumber \\ &&
\hspace*{1cm}
\left. +
4 ( \Delta \omega_k  t)^2 \right]^{1/2} < 1
\,.
\end{eqnarray}
It is readily proved that the potential correction $\Re \langle \delta V_{\chi} \rangle $
is always negative, implying the tachyon-like instability at all higher orders.

There are other types of perturbation.
Consider a successive decay, 
\begin{eqnarray}
&&
|2_s\rangle_{\chi} \rightarrow  |1_s\rangle_{\chi} \rightarrow |0_s\rangle_{\chi}
\,,
\end{eqnarray}
that occurs the same site $s$ starting from the initial state $|\Psi(2,0) \rangle $.
If the successive decay always occur, one can sum up in the expectation value
to derive
\begin{eqnarray}
&&
\sum_n  \overline{{\cal E}_{4n} }
= \frac{1}{1 - \overline{{\cal E}_4 }}
\,,
\\ &&
\overline{{\cal E}_4 } = \tilde{\chi}^4 
\left( 1 - e^{- 4i \Delta \omega_k t  } - 4i \Delta \omega_k t \right)
\,.
\end{eqnarray}
If two decay patterns occur simultaneously, one derives
\begin{eqnarray}
&&
\langle \Psi(2; 0)| U(t,0) | \Psi(2; 0) \rangle = 
\frac{1}{1 - {\cal E}_2^2 - \overline{{\cal E}_4 }}
\,,
\\ &&
{\cal E}_2^2 + \overline{{\cal E}_4 } = 
\tilde{\chi}^4
\left( (e^{- 2i \Delta \omega_k t }  -1 +  2i \Delta \omega_k t)^2
\right.
\nonumber \\ &&
\hspace*{1cm}
\left.
+ 1 - e^{-4i \Delta \omega_k t } - 4i \Delta \omega_k t \right)
\,,
\\ &&
\tilde{\chi}^4 = 2.931 \times 10^{-7} \, (c_A M_{\rm P})^4 (\frac{z_0 }{z })^{18}
(\frac{\chi}{M_{\rm P}})^4
\,.
\end{eqnarray}
Note the sign difference of two geometric series, and
the instability of identified potential correction holds everywhere for the two cases.
We call the type that contains $1/(1 - {\cal E}_2^2 - \overline{{\cal E}_4} ) $
type \lromn2 model, while the previous one type \lromn1 model.
In type \lromn2 model tachyon instability starts from the term,
$\propto - \chi^4$ at the field origin.
When we later discuss the inflaton exit problem, we shall compare
results of these two types.

A more general type of decay patterns are possible if
one starts from the initial state $|\Psi(n; 0) \rangle \,, n\geq 3$.
It is more reasonable to adopt higher type of models,
since inflation is likely to end with highly excited $\chi$ states.
For simplicity we consider type \lromn2 alone as a representative
of these general decay types and compare results in two cases of
\lromn1 and \lromn2 types.

Our higher order calculations in perturbation theory is summation of a special class
of diagrams, which we believe to be dominant over others.
A more explicit demonstration of this assertion is difficult, but
may be worthwhile to attempt.

The tachyon instability is a pre-requisite to destabilize 
the effective field potential, defined by  the potential correction 
$ \langle \delta V_{\chi} \rangle $ added to the original $V_{\chi}$.
We shall calculate the time of destabilization denoted by $t_{\rm D}$
defined by the time when $ \langle \delta V_{\chi} \rangle $ becomes
equal to the maximum value of $V_{\chi}$.
The condition gives
\begin{eqnarray}
&&
 \frac{4 e^{-2}}{\gamma_{\chi}^2} V_0 = 
\frac{M_{\rm P}^4 }{ 6} (\frac{z }{z_0 } )^6\, \Im {\cal E}_2
\,,
\end{eqnarray}
with $4 e^{-2} = 0.541$.
The favored value of the potential height $ V_0 = (10^{16}{\rm GeV})^4$ 
is suggested  in \cite{koy 24-1} from observations.
Thus, there exists a mismatch between the potential correction
of order $M_{\rm P}^4$ and the original potential $V_0$,
unless the destabilization occurs close to the thermalization time.
This suggests a possibility of inflaton exit out of trapped region
immediately after emergence of tachyon instability.

We add a comment to results so far derived.
In order to somewhat reinforce this correlation, we  next consider more general potential
which has a harmonic part near the field origin, $\chi \approx 0$,
but its deviation from HO is made to appear far away from the field origin.
The inflaton field in the interaction picture  has time dependence of the form,
$\chi^{\pm } (t) e^{ \mp i m_{\chi} t}$.
One cannot explicitly time-integrate the interaction hamiltonian in this case.
We shall focus on the potential region close to HO regime, but allow its small deviation
from HO.
The approximation of slow time variation may then be useful taking expansion to the first derivative term,
\begin{eqnarray}
&&
\chi(t') \approx \chi(t) + \dot{\chi} (t)\, (t'-t)
\,,
\label {approximate field variation}
\end{eqnarray}
near $t'= t$ (taken as a reference point of time). 
This makes it possible to integrate both terms 
$\propto \chi(t) $ and $\propto  \dot{\chi}(t)$ over $t'$;
\begin{eqnarray}
&&
-i \int_0^{t} dt'\, H'_{\rm int} \sim - i\,  2\sqrt{2} 
\frac{ c_A \Omega_k m_{\chi}}{ 2 \Delta \omega_k} 
\nonumber \\ &&
\left( 
- (a_k^{\dagger} a_{-k}^{\dagger} + a_k a_{-k} ) \left( \chi(t)\, \sin 2 \Delta \omega_k  t
\;\;
\right.
\right.
\nonumber \\ &&
\left.
- \frac{ \dot{\chi}(t)}{ 2 \Delta \omega_k}\, (1 - \cos 2 \Delta \omega_k  t 
+ 2 \Delta \omega_k 
t \sin 2 \Delta \omega_k  t )
\right)
\nonumber \\ &&
 + \left( i (a_k^{\dagger} a_{-k}^{\dagger} - a_k a_{-k} ) \left( \chi(t)( 1 - \cos 2\Delta \omega_k  t)
\right.
\right.
\nonumber \\ &&
\left.
- \dot{\chi}(t)\,  t \cos 2\Delta \omega_k  t
\right)
\,.
\end{eqnarray}
The on-shell limit of this integral is 
\begin{eqnarray}
&&
-i \int_0^{ t} dt'\, H'_{\rm int}(t')
 \rightarrow  i\,  2\sqrt{2} \frac{ c_A m_{\chi} \Omega_k } { 2 \Delta \omega_k }  
\nonumber \\ &&
2 \Delta \omega_k  t\,
\left(a_k^{\dagger} a_{-k}^{\dagger}  (
\chi +   \frac{\dot{\chi}}{ 2 \Delta \omega_k} \, 3\Delta \omega_k  t ) 
\right.
\nonumber \\ &&
\hspace*{-0.5cm}
\left.
-  i \, a_k^{\dagger} a_{-k}^{\dagger}  \left(  
\chi \Delta \omega_k t + \dot{\chi}  t   \right) + ({\rm h.c.}) \right)
+ O(\,(\Delta \Omega_k  t)^3 \,)
\,. 
\end{eqnarray}
New terms $\propto \dot{\chi}$ added to the HO case enhances the unstable real part effect if the sign condition $\chi \dot{\chi} \, \Delta \omega_k > 0 $ holds.

Throughout discussions so far we neglected effect of Hubble friction. 
This is justified when instability and particle production occurs
much faster than the Hubble time, as is the case.
When one considers a long time behavior, this should be corrected,
as is made in the equation (\ref{ho field eq with hubble}).

\section
{Super-radiant stage of preheating}

We consider the preheating stage in which inflaton  $\chi$ field equation
is described by HO (harmonic oscillator) potential.
After recapitulating the Dicke model, we shall show that
our quantum model provides extensions of the Dicke idea in two directions;
extended coherence area and more general level structures.

\subsection
{Dicke algebraic model and its identification to our quantum theory}

There exists an important relation on correlated changes of the numbers
of inflaton quanta and radiation pairs.
Each time the inflaton decays, it produces a radiation pair,
hence the total number of inflaton and radiation pair is conserved:
$\Delta ( n_{\chi} + n_{\rm rad-pair} ) =0$.

At a single site it is assumed that there are only two levels, the ground state $|0\rangle $
and the excited state $|1 \rangle$, and the excited state is allowed to pair emit along with
the pair annihilation elevating the ground state to the excited state.
The spin algebra of two-level system at a single site is defined by
\begin{eqnarray}
&&
\hspace*{-0.3cm}
j_1 = \frac{1}{2} ( |1\rangle \langle 0| + | 0\rangle \langle 1 | )
\,, \hspace{0.3cm}
j_2 = \frac{i}{2} (- |1\rangle \langle 0| + | 0\rangle \langle 1 | )
\,, 
\\ &&
j_3 = \frac{1}{2} ( |1\rangle \langle 1| - | 0\rangle \langle 0 | )
\,.
\end{eqnarray}
They satisfy the SU(2) commutation relation,
\begin{eqnarray}
&&
[j_k \,, j_l ] = i \epsilon_{klm} \, j_m
\,.
\end{eqnarray}

Dicke introduces the ansatz that the correlated change proceeds
by a hypothetical angular momentum change, thereby
maintaining phase coherence and permutation symmetry of
site exchange.
The spin algebra is extended to the angular momentum algebra consisting of many 
spins at different sites $s$,
\begin{eqnarray}
&&
\vec{J} = \sum_s \vec{j}^{(s)}
\,, \hspace{0.3cm}
[\vec{j}^{(s)} \,, \vec{j}^{(s')}] = 0 \,; \; \; {\rm if \;} s \neq s'
\,.
\end{eqnarray}
The basic Dicke assumption is that photon emission occurs via $J_-$ and
photon absorption via $J_+$, and the rate of photon emission is in proportion to
\begin{eqnarray}
&&
(J_-)^{\dagger} J_- = J_+ J_- = \sum_{s s'} j_+^{(s')} \, j_-^{(s) }
\,.
\end{eqnarray}
The second Dicke's assumption concerns the initial state, for which it is taken all sites
in the excited state, hence 
\begin{eqnarray}
&&
|\Psi \rangle_{\chi} = \Pi_s\, | 1\rangle_s
\,.
\label {dicke initial state}
\end{eqnarray}
The projected angular momentum $M$ is defined in a convenient range,
$M = N_s, N_s -1, N_2-2, \cdots 0$ such that it is always non-negative.
The relation $ M + n = N_s$ holds.

We now come to the important correspondence relation.
We may identify the angular momentum operator of Dicke type to
our quantum operator according to
\begin{eqnarray}
&&
\hspace*{1cm}
J_{\mp} \; \Leftrightarrow \; Q^{(\pm)}(k)
\,, 
 \\ &&
 |M \rangle 
= \left( Q^{(+)}(k) \right)^{N_s -M }\, \Pi_s \, |1_s \rangle\!_{\chi}  | 0 \rangle\!_{\rm rad}
\,,
\label {operator identification}
\end{eqnarray}
where $|1_s \rangle\!_{\chi} $ stands for the $\chi$ first excited state at site $s$.
The site permutation symmetry in the Dicke model is automatically respected
in our quantum model due to the site sum in  $Q^{(+)}(k)$.

Extensions beyond the Dicke model
are readily taken into account in our model.
If we assume that all sites are occupied in the first excited state of $\chi$ quantum $|1_s \rangle\!_{\chi}$, the initial state is given by 
$|\Psi\rangle_i^{(1)} = |\Psi(1; 0) \rangle =\Pi_s \, |1_s \rangle\!_{\chi}  | 0 \rangle\!_{\rm rad} $.
This choice is inevitable in the Dicke model that assumes two atomic levels.
We have seen that another choice of $|\Psi(2; 0) \rangle $ adds a new interesting feature
in our model.
This choice provides a departure from the Dicke model.
We call these Dicke-type \lromn1, Dicke-type \lromn2 models in the present work.

\subsection
{Macro-coherent amplification}

Our model hamiltonian gives eJBD scalar decay into a pair of gauge bosons with
momentum correlation, while the original Dicke model is based on a single photon decay
of excited atoms.
The phase coherence necessary for the enhanced decay is limited
in the Dicke model to the wavelength $\lambda$
region, since the phase factor $e^{i 2\pi x/\lambda}$ may deviate
outside of $|x| > \lambda/2\pi$ from $e^{i 2\pi/\lambda} \sim 1$.
The situation dramatically changes for decay of correlated gauge boson pair decay,
since this time the phase factor  relation $e^{i (\vec{k} - \vec{k} ) \cdot \vec{x}} = 1$ 
holds beyond the wavelength region.
This is the essence of macro-coherent enhancement \cite{macro-coherence},
which was experimentally verified in para-H$^2$ molecule into
two photons, a weak QED process \cite{two-photon macro-coherence}.

The number of inflaton quanta (equal to the site number $N_s$) that participate in macro-coherence
is the $\chi$ number density $\rho_{\chi}/m_{\chi}$ times the Horizon volume,
\begin{eqnarray}
&&
N_s = \frac{\rho_{\chi}}{m_{\chi}} H^{-3} = 3^{3/2} \, \frac{\rho_{\chi} M_{\rm P}^3}{m_{\chi}^4 \chi^3} 
\, (1 + r )^{-3/2}
\,,
\end{eqnarray}
with $r$ the ratio of $\chi$ energy density to the radiation energy density.
At the Planck epoch all quantities here are of order, the Planck values.
Hence this number is of order unity.
As the preheating proceeds, $N_s$ dramatically increases.
At the end of preheating one may use 
the relation $m_{\chi}^2 \chi^2 \sim V_0 \approx (10^{16} {\rm GeV})^4$,
hence $m_{\chi} \sim 10^{12}\,$GeV, giving $N_s = O(10^{12})$.
In intermediate times one may use the adiabatic relation, a constant  $a(t) T(t) $
independent of time $t$,
which gives the redshift ratio $z_R/z_0 \sim 10^{-2} $ at thermalization to at Planck epoch.
This leads to $N_s$ time evolution,
\begin{eqnarray}
&&
N_s = O(1) (\frac{M_{\rm P}}{m_{\chi}})^2 \frac{M_{\rm P}}{\chi} =O(1) (\frac{z_0}{z})^6 \frac{M_{\rm P}}{\chi}
\,.
\end{eqnarray}

There is thus a considerable ambiguity in estimation of the site number $N_s$.
In view of this ambiguity we leave this number unspecified except when quantified numbers are given.

\begin{figure*}[htbp]
\begin{center}
\includegraphics[width=0.6\textwidth]{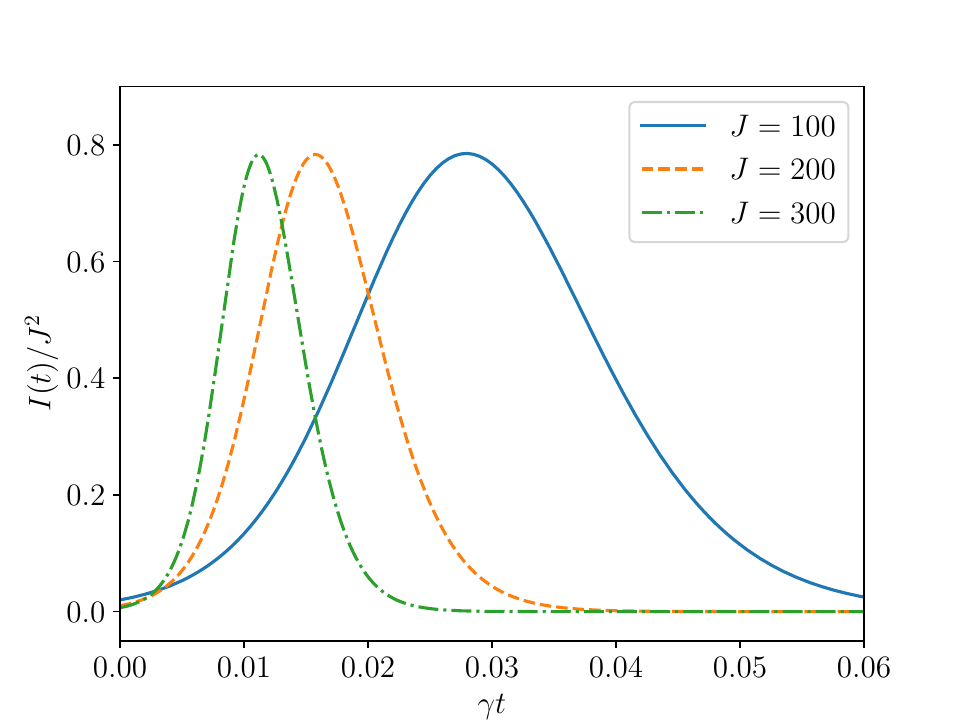}
  \caption{
Energy flux given by (\ref{energy flux eq}) in our macro-coherent  Dicke model:
three cases of different site numbers given by $ 2 J+1$ are plotted
against perturbative decay rate $\times$ time.
The peak positions follow the delayed time
 formula given by  $\propto \ln (2J+1)/(2J+1)$ given in (\ref{delayed time}).
}
   \label {energy flux}
 \end{center} 
\end{figure*}

Cosmic scale factor $a(t)$ in different preheating stages is calculated 
using the energy sum of 
eJBD quanta  and produced gauge boson pairs;
\begin{eqnarray}
&&
(\frac{\dot{a}}{a})^2 = \frac{1 }{ 3 M_{\rm P}^2 } \left( \rho_{\chi} + \rho_{\rm rad}
\right)
\,,
\\ &&
\rho_{\chi} = \frac{1}{2} \dot{\chi}^2 + \frac{1}{2} m_{\chi}^2 \chi^2 \propto a^{-3}
\,, \hspace{0.3cm}
\rho_{\rm rad} \propto a^{-4}
\,.
\end{eqnarray}
Contribution from interaction energy has been neglected.
In the initial stage of inflaton dominance one can ignore radiation term $\propto \rho_{\rm rad}$,
and one derives the time dependence of matter-dominated universe,
$a(t) \propto t^{2/3}$, while in later stage of radiation-dominance the scale factor varies
$a(t) \propto t^{1/2}$.
They give the power law of Hubble rate, $H(t) = 2/(3t)$ or $1/(2t)$, respectively.

Decay process of a collected body of excited states goes on
in an entirely different way from stochastic exponential decay law.
Coherence is first developed, along with slowly proceeding exponential decay,
but with accumulated coherence a collected body of excited states suddenly disappear, 
leaving behind the ground state.
The algebraic equation that describes the coherence evolution was
proposed by Dicke: 
his equation of occupancy probability $P_M(t)$ at the projected angular momentum $M$ is
\begin{eqnarray}
&&
\frac{d P_M}{\Gamma_k^R\, dt}  =   |\langle J, M+1 | J^+ | J, M \rangle |^2
P_{M+1}
\nonumber \\ &&
 -   |\langle J, M-1 | J^- | J, M \rangle |^2 P_{M}
\,. \hspace{0.3cm}
\Gamma_k^R = \Gamma_B N_s
\label {sr rate eq}
\end{eqnarray}
supplemented by well-known relations,
$ |\langle J, M+1 | J^+ | J, M \rangle |^2 = (N_s - M) ( N_s + M +1)$ etc.
The quantity $\Gamma_k^R$ is decay constant enhanced by the total number of sites
within the horizon.
One can prove the law of probability conservation from (\ref{sr rate eq}),
\begin{eqnarray}
&&
\frac{d}{dt}\, \sum_{M= -J}^{J} P_M = 0
\,.
\end{eqnarray}

Delayed decay time denoted by $t_d$ is the important concept to characterize the sudden collapse of
excited states \cite{sr textbook}, and is given by
\begin{eqnarray}
&&
t_d = \frac{\ln N_s }{N_s }\, \frac{1}{\Gamma_k^R } 
\,.
\label {delayed time}
\end{eqnarray}
Note that the combination of denominator $N_s \Gamma_k^R = \Gamma_B N_s^2$ is doubly
enhanced by the site number $N_s$.
The depletion of excited states leaves behind $\chi$ field corresponding to
zero-point oscillation.
We may call the factor $N_s^2/\ln N_s$ the speed up factor
of macro-coherent super-radiance.

In Fig(\ref {energy flux}) we illustrate time evolution of energy flux $I(t)$
given by 
\begin{eqnarray}
&&
\frac{I(t)}{J^2} = 
- \frac{1}{J^2 } \sum_{M= - J \sim J} \, M \frac{d P_M(t)}{dt}
\,,
\label {energy flux eq}
\end{eqnarray}
in our macro-coherent Dicke model.
As the number of participating sites increases, the peak of the flux
shifts to shorter times approximately according to the formula
of delayed time $\propto \ln (2J+1)/(2J+1)$ given in (\ref{delayed time}),
at the same time with narrower widths of duration time,
as expected.
Perturbative decays occur only for $1- e^{-0.06} \sim 0.058$ 
(at the time of $0.06 \times$ the usual lifetime) of participating inflatons,
while the decay is completed (almost completed in the $J=100 $ case) 
in three $J$ cases shown.

\section
{Transition from inflationary  to dark energy stages of acceleration}

\subsection
{Inflaton escape probability}

We first explain our method of solving the exit problem.
In our quantum formulation we can set up the Schroedinger equation
of homogeneous inflaton field $\chi$;
\begin{eqnarray}
&&
i \hbar \frac{\partial \Psi}{\partial t} = - \frac{\hbar^2}{2} 
\frac{\partial^2 \Psi}{\partial \chi^2}
+ \tilde{V}(\chi) \Psi
\,, 
\\ &&
\tilde{V}(\chi) =  \left( V_{\chi} + \Re \langle \delta V_{\chi} \rangle \right)\,
H^{-3}
\,.
\end{eqnarray}
This equation is solved under the WKB ansatz in $\hbar \rightarrow 0$ limit;
\begin{eqnarray}
&&
\Psi_E (\chi, t) = \exp[-  \frac{i \, E t}{\hbar} + \frac{i}{\hbar} W(\chi; E) ]
\,.
\end{eqnarray}
To leading orders, its solution is given by
\begin{eqnarray}
&&
\Psi_E (\chi, t) = N\, (E- \tilde{V}(\chi)\,)^{-1/4}\, \exp[- i E t] 
\nonumber \\ &&
\hspace*{0.5cm}
\sin \left( \int_a^{\chi}d\chi \sqrt{E-\tilde{V}(\chi)}
- \frac{\pi}{4} \right)
\,,
\end{eqnarray}
with $N$ the normalization and with $\hbar$ dependence deleted.
$a$ is the turning point at which $E-\tilde{V}(\chi = a) = 0$.
The reflected component is obtained by the replacement,
\begin{eqnarray}
&&
\sin \left( \int_a^{\chi}d\chi \sqrt{E-\tilde{V}(\chi)}
- \frac{\pi}{4} \right) \rightarrow 
\nonumber \\ &&
\frac{\exp[i \left( \int_a^{\chi}d\chi \sqrt{E-\tilde{V}(\chi)} - \frac{\pi}{4} \right)]}{2i}
\,.
\end{eqnarray}

The potential correction given to the second order of interaction
describes its behavior near the field origin, establishing the tachyon mass.
But it is not sufficient to describe the potential region around
the original potential maximum.
The potential correction should  be applied to larger field values, and
we adopt type \lromn2 model giving the potential correction 
$\propto \ln (1- {\cal E}_2 - \overline{{\cal E}}_4)$, as derived 
in (\ref{potential expectation-value rel}).

The inflaton distribution probability, using the reflected component, is given by
\begin{eqnarray}
&&
| \Psi_E(\chi, t)|^2 = 
\frac{ |N|^2 }{ 4 \sqrt{| E- H^{-3} \langle \delta V_{\chi} \rangle |}}
\,, \hspace{0.3cm}
| E- H^{-3} \langle \delta V_{\chi} \rangle | 
\nonumber \\ &&
\hspace*{1cm}
= |\, E -   \frac{\sqrt{3}}{2} M_{\rm P} 
(\frac{z }{ z_0})^{6}
\Im \ln (1 - {\cal E}_2^2 - \overline{{\cal E}}_4)\, |
\,,
\label {wkb type-2 probability}
\\ &&
\hspace*{1cm}
|{\cal E}_2^2 + \overline{{\cal E}}_4 | < 1
\,.
\label {type-2 constraint}
\end{eqnarray}
The constraint (\ref{type-2 constraint}) gives a time dependent upper limit
on $\chi$ field;
\begin{eqnarray}
&&
\hspace*{1cm}
\frac{\chi}{M_{\rm P}} < \frac{32\pi^2}{3 \sqrt{6}} 
\frac{1}{c_A M_{\rm P} } (\frac{z }{ z_0})^{9/2}\, \times
\nonumber \\ &&
\hspace*{-0.5cm}
\left[ |( 1 - e^{- 2i \Delta \omega_k t } - 2i \Delta \omega_k t)^2 +1 
- e^{-4i \Delta \omega_k t} -4i \Delta \omega_k t  |\right]^{-1/4} 
\,.
\nonumber \\ &&
\end{eqnarray}

\begin{figure*}[htbp]
 \begin{center}
\includegraphics[width=0.6\textwidth]{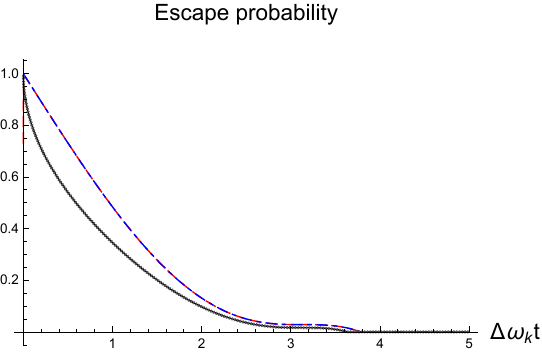}
   \caption{
Escape probability, calculated using WKB probability formula (\ref{wkb type-2 probability})
of type \lromn2 and \lromn1 models with the constraint (\ref{type-2 constraint}),
 of eJBD inflaton out of trapped region plotted against
a small  dimensionless time $\Delta \omega_k t $
(in terms of $m_{\chi}$ the time axis should be multipled 
by $m_{\chi}/\Delta m_{\chi} = 10 \sim 100$).
The trapped region is defined by the interior to the potential maximum
of the original potential $V_{\chi}=1.8 \times 10^{-13} $ in the Planck unit
at $\chi_{\rm max}$.
The inflaton energy is taken as $E=0 $ and to avoid the
singularity the region inside the trapped region is defined by 
$\chi_{\rm min} = \chi_{\rm max}/100$.
Used parameters are a common $\gamma_g=1$
and $z/z_0 = 0.1$ of \lromn2 type in solid and 
$0.01$ dotted black,
$0.1$ of \lromn1 type in dashed orange and $0.01 $ dash-dotted blue.
}
   \label {escape probability 2}
 \end{center} 
\end{figure*}

One must be careful about the artifact of WKB approximation which
gives a divergence at the classical turning point at
$ E- H^{-3} \langle \delta V_{\chi} \rangle = 0$.
We avoid the problem by an appropriate choice of the energy $E$.
The escape probability differs in different models,
for instance, type \lromn1 or type \lromn2 model.
We illustrate four parameter cases of escape probability
$| \Psi_E(\chi, t)|^2 $  in Fig(\ref{escape probability 2})
for $E=0$.
The normalization $N$ has been calculated including
the interior trapped region.
Type \lromn1 model gives a larger escape probability than type \lromn2 model.

The termination time of escape is estimated by the time
when $\chi$ energy is drained into radiation.
One method of estimation is to calculate
the total decay rate and  use the constraint relation,
${\cal E}_2=1$ or ${\cal E}_2^2 + \overline{{\cal E}}_4=1$, depending on
the model type, \lromn1 or \lromn2.
Total decay rate in type \lromn1 model is given by
\begin{eqnarray}
&&
\hspace*{1cm}
\Gamma_B 
\sum_n |{\cal E}_2 |^{2n} =\frac{\Gamma_B }{1 - |{\cal E}_2 |^2} 
\label {inflaton decay rate}
\,,
\\ &&
|{\cal E}_2 |^2 = 4 \tilde{\chi}^2 |\,  1 - 2 \cos 2 \Delta \omega_k t +
2\Delta \omega_k ( \Delta \omega_k - \sin2 \Delta \omega_k t  \,|
\,.
\nonumber \\ &&
\end{eqnarray}

\begin{figure*}[htbp]
 \begin{center}
\includegraphics[width=0.6\textwidth]{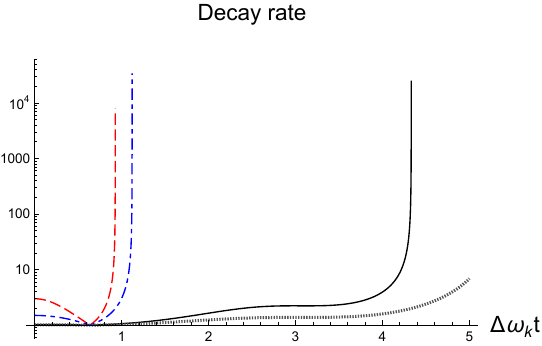}
   \caption{
Total decay rate given by $1/(1- | {\cal E}_2|^2)$: 
Four cases of $(c_A M_{\rm P}), z, \chi/M_{\rm P}$ in type \lromn1 are plotted:
(0.01,0.1,0.1) in solid
black, (0.01,0.05,0.1) in dashed orange (0.01,0.05, 0.05) in dash-dotted blue, and (0.01,0.1,0.05) in dotted black.
}
   \label {escape end-point}
 \end{center} 
\end{figure*}

The function $1/( 1 - |{\cal E}_2 |^2) $ that appears in (\ref{inflaton decay rate})
 is illustrated in Fig(\ref{escape end-point}),
which shows that the escape should be completed within
a few to several times $\Delta \omega_k t$ at relevant redshifts $z_{\rm escape} $.
Combined with redshift values used in Fig(\ref{escape end-point}),
 we observe that
the escape occurs early during the preheating stage.

The potential correction is forced to disappear when all inflaton excitation
energy is drained into radiation pairs.
The drained time denoted by $t_D$ or its redshift $z_D$
is estimated by equating HO $\chi$ energy density $\rho_{\chi}$
(its non-decay part) $\propto z^3$
to the maximum potential energy density  $2V_0 /\gamma_{\chi}^2$.
Using our favorite values of $V_0\,, \gamma_{\chi} $, we derive 
\begin{eqnarray}
&&
\frac{z_D}{z_0} = \left(\frac{2V_0} {\gamma_{\chi}^2\, M_{\rm P}^4} \right)^{1/3} 
\sim 2.1 \times 10^{-3}
\,.
\label {drained chi time}
\end{eqnarray}
This is roughly of the same order of magnitude as the thermalization time,
and  is  later than
the escape time of order $z_{\rm escape}/z_0 = O(0.1 \sim 0.01)$.

\subsection
{Epochs of thermalization and potential restoration}

We have so  far considered transverse gauge boson pair production at the preheating stage.
This process creates a universe dominated by gauge boson pairs far from chemical and thermal equilibrium.
The minimum requirement for thermalization is to produce energetically different
particles and utilize scattering between particles of different energies,
thereby exchanging their energies.
The simplest mechanism is to take into account massive gauge boson decay
into lighter fermion anti-fermion pair.
Recall that prior to thermalization the gauge symmetry is spontaneously broken
and various gauge bosons are massive and they can decay into lighter fermion pairs.

In \cite{koy 24-2} we estimated the time at which thermalization
takes place, and the result may be given in terms of the redshift factor,
\begin{eqnarray}
&&
\frac{z_{\rm th}}{z_0} = O(2 \times 10^{-3} \sim 4 \times 10^{-2})
\,,
\label {thermalization z}
\end{eqnarray}
using coupling constants and the number of particle species in
standard and grand unified models.

The redshift ratio (\ref{thermalization z}) is larger 
than the ratio (\ref{drained chi time}), which means that
the termination of potential correction occurs after thermalization.
Without potential correction the original inflaton potential is restored,
and $\chi$ field continues to roll down towards the field infinity
in thermalized radiation-dominated universe.
The field motion describes a quintessential dark energy.
After thermal equilibrium is realized, there is an additional contribution
to the field potential which restores the standard model gauge symmetry.

One can set up an approximate set of time evolution equations for
eJBD field, radiation energy  density and the cosmic scale factor
(or the redshift factor).
But a more elaborate approach that takes into account thermalization
effect is desirable to derive conclusive results, since
thermalization and the  recovery to the original potential proceeds simultaneously.

\section
{Summary and outlook}

We have proposed a quantum theory of 
how the universe after inflation underwent the transition to
hot big band universe that realizes a kind of phase transition 
to the late-time accelerating universe.
Incorporation of particle physics into an extended class of Jordan-Brans-Dicke gravity
is of central importance, and we generalized the original JBD idea of
matter coupling maintaining  five independent gauge invariant terms
separately with different conformal factors.
We determined the major matter coupling of eJBD inflaton to various gauge-boson pairs
by the unambiguous trace anomaly formula.

The eJBD field coupling via the trace anomaly,
giving rise to gauge-boson pair decay, is of fundamental importance
in enhancing super-radiant inflaton decay due to a high level of 
macro-coherence.

In the initial preheating stage the inflaton motion is described by a harmonic oscillator,
and interaction of inflaton to matter induces 
the inflaton decay into gauge bosons of correlated momentum pair, back-to-back.
Quantum theory of inflaton and gauge-boson pairs have been fully developed.
Standard particle physics techniques lead to an orchestrated particle pair production
accompanied by tachyonic instability giving rise to an inflaton potential change.
The tachyon instability gives  a temporary epoch of destabilized inflaton potential,
which pushes the inflaton trapped in the harmonic well to cross over a local potential maximum, ultimately towards the field infinity.
This offers a conversion mechanism  that bridges between inflationary
acceleration and the dark energy acceleration.
The present single eJBD field model is more economic, hence more restrictive than
the model  of two fields advocated in our previous work \cite{koy 24-1}.

Our new theory is a tremendous simplification of calculation over
the previous methods \cite{kls}, \cite{my 95 and fkyy} that
 relied on mode summation in instability bands of the Mathieu chart.
The new approach  bypasses this complicated procedure,
yet one can derive  important issues connected to
the Dicke super-radiance model. 
Two extended types of Dicke model have been considered in detail.

Type \lromn1 Dicke model gives slightly larger escape probability
and faster ending than type \lromn2,
but  not overwhelmingly better.
Both types are acceptable to give large enough escape probabilities.

Convenient tools of dedicated numerical computations need to be developed, but
it should be clear that the scenario presented in this work
is realized using an acceptable set of parameters in the model.

Our model is described by a few parameters of eJBD gravity related to inflaton coupling of standard matter.
These parameters greatly influence the late stage of cosmic evolution.
As is made evident in \cite{koy 24-1}, the effective scalar mass after nucleo-synthesis
is of the order of the inverse Hubble size, hence in problems of astrophysical interest,
the scalar is essentially massless with its coupling to matter of the gravitational strength.
This may or may not influence gravitational wave emission at merger events.
It is expected that mergers of black hole and neutron star are a good target
to search for deviation from the prediction of general relativity
\cite{koy bh-ns merger}.
Thus, our model can be tested in forthcoming observations and laboratory
experiments.
These problems shall be worked out in separate publications.

\section
{Appendix: Relevance of trace anomaly}

\begin{figure*}[htbp]
 \begin{center}
\includegraphics[width=0.6\textwidth]{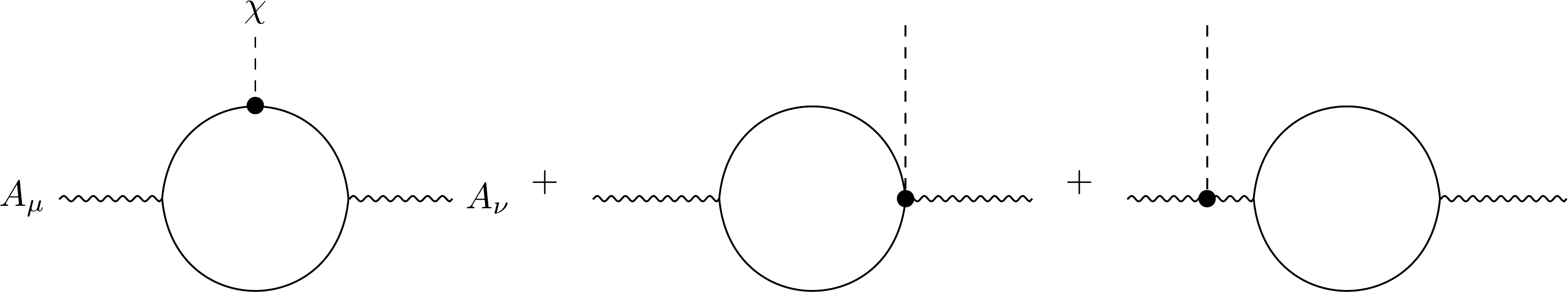}
   \caption{
Three diagrams contributing to inflaton decay into massless gauge boson pair.
Solid lines are for fermions  attached to wavy gauge bosons and
long dotted inflaton field, while black circles are the trace anomaly. 
}
   \label {anomaly generated pair decay}
 \end{center} 
\end{figure*}

We first discuss how a new conformal coupling
 term  should be added to those of our previous works \cite{koy 24-1},
\cite{koy 24-2}.

To see how the eJBD field couples to the trace of the energy-momentum tensor, we begin by writing the action as
\begin{eqnarray}
&&
\hspace*{-0.3cm}
    S =
    \int d^4x \sqrt{-g_J}\left[
        -\frac{M_P^2}{2}F_g(\chi) R_J + {\cal L}_{\chi} + {\cal L}_{\rm SM}
    \right],
\end{eqnarray}
where ${\cal L}_{\chi}$ and ${\cal L}_{\rm SM}$ are assumed to include couplings of the eJBD field 
in a gauge invariant manner as in \cite{koy 23}.
To transform from the Jordan metric frame to the Einstein metric frame, we perform the Weyl rescaling given by $g_{J\mu\nu} = F^{-1}_g(\chi)g_{\mu\nu} $.
Suppose that $\chi$ takes a nonzero homogeneous contribution $\chi=\chi_0=$ constant and that we redefine $\chi$ as a fluctuation on $\chi_0$, namely, $\chi\to\chi_0+\chi$ with $|\chi/\chi_0|\ll 1$.
This setup allow us to expand as
\begin{eqnarray}
&&
    g_{J\mu\nu} = F^{-1}_g(\chi_0)\left(
        1 - \partial_\chi \ln F_g(\chi_0) \chi + \cdots
    \right)g_{\mu\nu}.
\nonumber \\ &&
    \equiv
    F^{-1}_g(\chi_0)(g_{\mu\nu}+\delta g_{\mu\nu}).
\end{eqnarray}
Note that the overall constant factor may be removed by a simple rescaling, so we omit the factor $F_g^{-1}(\chi_0)$ in the following argument.

The relevant interaction of $\chi$ can be identified by varying 
the action $S$ with respect to 
$\delta g_{\mu\nu}$, to derive
\begin{eqnarray}
&&
    S_{\rm int} = \frac{1}{2}\int d^4x \sqrt{-g}T_{\mu\nu}\delta g^{\mu\nu}
\nonumber \\ &&
= -\int d^4x \sqrt{-g} \gamma_J \frac{\chi}{2M_P} T
\,,
\end{eqnarray}
with $ T = g^{\mu\nu}T_{\mu\nu}$.

There is a piece of contribution to the trace of the  energy-momentum tensor in
the standard model,
\begin{eqnarray}
&&
    T_{\mu}^{\mu} = \sum_{A_\mu}\frac{4-d}{4} F^{\mu\nu}F_{\mu\nu} + (\cdots)
\,.
\label {trace from squared field strength}
\end{eqnarray}
The trace anomaly exists since the gauge boson pair contributes at one loop level.
We find by computing the self-energy diagram of gauge boson that
this gives rise to a finite contribution to the
decay amplitude, the divergent contribution $\propto 1/(4-d)$ 
in dimensional regularization scheme canceling $4-d$
in (\ref{trace from squared field strength});
\begin{eqnarray}
&&
  -\frac{b_A\alpha_A}{8\pi} F^{{\rm A}\mu\nu}F^{\rm A}_{\mu\nu}
\,, \hspace{0.3cm}
 \frac{d\alpha_A^{-1}}{d\ln\mu} = -\frac{b_A}{2\pi}
\,.
\end{eqnarray}
$b_A$ thus defined is the beta-function coefficient,
with $\mu$ being the renormalization scale.

\vspace{0.5cm}
\begin{acknowledgments}
This research was partially
 supported by JSPS KAKENHI  Nos. 21H01107 (KO) and 21K03575(MY).

\end{acknowledgments}

\end{document}